\documentclass[runningheads]{llncs}

\usepackage{url}
\usepackage{svg}
\usepackage{amsmath,amsfonts}
\usepackage{amssymb}
\usepackage{graphicx}
\usepackage{pgfplots}
\usepackage{tikz}
\usepackage{listings}
\usepackage{color}
\usepackage[dvipsnames]{xcolor}
\definecolor{dkgreen}{rgb}{0,0.6,0}
\definecolor{gray}{rgb}{0.5,0.5,0.5}
\definecolor{mauve}{rgb}{0.58,0,0.82}
\usepackage[ruled,vlined,linesnumbered,noend]{algorithm2e}
\usepackage[T1]{fontenc}
\usepackage{tabularx}
\usepackage{multirow}
\usepackage{tcolorbox} 
\usepackage{booktabs}
\usepackage{float}
\usepackage{algorithmic}
\algsetup{linenosize=\scriptsize}
\usepackage{setspace}
\usepackage{array}
\usepackage{textcomp}
\usepackage{stfloats}
\usepackage{tcolorbox}
\usepackage{verbatim}
\usepackage{tabularx}
\usepackage{makecell}
\usepackage{bbding}
\usepackage{graphicx}
\usepackage{balance}
\usepackage{threeparttable}
\usepackage{pgfplots}
\usepackage{tikz}
\usepackage{caption}
\usepackage[final]{pdfpages}
\usepackage[commandnameprefix=always, defaultcolor=red]{changes} 
\usepackage[labelformat=simple]{subcaption}

\usepackage{booktabs,multirow,bm,wrapfig,mathtools,dashbox,pifont,soul,enumitem}
\usepackage[commandnameprefix=always, defaultcolor=red]{changes} 

\usetikzlibrary{decorations.pathmorphing}   
\usetikzlibrary{fit}                        
\usetikzlibrary{backgrounds}                
\usetikzlibrary{calc}
\usetikzlibrary{positioning,shadows,arrows}
\usetikzlibrary{shapes,shapes.multipart, shapes.geometric}

\RequirePackage{mdframed}
\renewmdenv[leftmargin=0.4em, rightmargin=0.4em%
innerleftmargin=0.6em,innerrightmargin=0.6em]{quote}

\newcommand{\wc}[1]{\textcolor{blue}{\textbf{[Wen: }\textsf{#1}\textbf{]}}}

\newcommand{\aero}{\textsc{Aero}\textsc{Req2LTL}\xspace}

\newcommand{\spacekg}{\textsc{SpaceKG}\xspace}
\newcommand{\spacerdl}{\textsc{SpaceRDL}\xspace}

\setlist[itemize]{leftmargin=*}
\setlist[enumerate]{leftmargin=*}
\newlist{steps}{enumerate}{1}
\setlist[steps, 1]{label = \textbf{RQ\arabic*.}}

\usepackage{hyperref}
\usepackage{url}
\usepackage{xcolor,colortbl,color}
\hypersetup{
 colorlinks=true,
 linkcolor=purple,
 citecolor=violet,
 filecolor=magenta,      
 urlcolor=cyan,
}

\newcommand{\eg}{\hbox{\textit{e.g.}}\xspace}

\setlist[itemize]{leftmargin=*}
\setlist[enumerate]{leftmargin=*}

\begin{document}
\title{Automated LTL Specification Generation from Industrial Aerospace Requirements}

\author{Zhi Ma\inst{1} \and Xiao Liang\inst{1} \and Cheng Wen\inst{2,}\thanks{Corresponding authors: Cheng Wen and Bin Gu.\\ mazhi@xidian.edu.cn, wencheng@xidian.edu.cn.} \and Rui Chen\inst{3,4} \and Bin Gu\inst{3,\star} \and \\
Shengchao Qin\inst{1,2} \and Cong Tian\inst{1} \and Mengfei Yang\inst{4}}
\institute{
School of Computer Science and Technology, Xidian University, Xi'an, China \and
Guangzhou Institute of Technology, Xidian University, Guangzhou, China \and
Beijing Institute of Control Engineering, Beijing, China \and
China Academy of Space Technology, Beijing, China \and
Beijing Sunwise Information Technology Ltd, China \\
}

\titlerunning{Automated LTL Generation for Aerospace Systems}
\authorrunning{Ma, Liang, Wen, Chen, Gu, Qin, Tian, Yang Meng-Fei}
\maketitle

\begin{abstract}
In the development and verification of safety-critical aero-space software, Linear Temporal Logic (LTL) has been widely used to specify complex system properties derived from requirements.
However, a significant gap remains in industrial practice: translating natural language (NL) requirements into formal LTL properties is a labor-intensive and error-prone process that requires rare expertise in both aerospace control engineering and formal methods.
While recent NL-to-LTL tools (\textit{e.g.}, NL2SPEC, NL2TL, NL2LTL) are capable of automating parts of this process, they often fail on real requirement documents in industrial settings, due to complex domain terminology or implicit temporal and logical structure.
To address these challenges, we present \aero, a framework that automates LTL property generation for aerospace requirements using large language models (LLMs), with two key industrial innovations: (i) a data dictionary that normalizes technical jargon into precise atomic propositions; and (ii) a template-based requirement language that makes temporal cues and logical relations explicit before translation. 
On a real aerospace dataset, \aero achieves 85\% precision and 88\% recall in LTL generation, and its outputs can be directly consumed by existing verification tools.


\end{abstract}

\section{Introduction}\label{sec:intro}
Software verification for safety-critical aerospace control software must provide strong and auditable assurance under tight engineering constraints, including frequent requirement changes, limited verification expertise, and strict expectations on evidence quality. 
In such settings, formal methods are often adopted because they can offer mathematically grounded guarantees beyond what traditional testing can typically provide. 
Within this domain, Linear Temporal Logic (LTL)~\cite{rozier2011linear} has emerged as the preferred formal specification language for expressing complex system properties.
Because aerospace software is inherently reactive—interacting continuously with an environment over time—its correctness depends not only on functional behavior but also on precise temporal properties~\cite{merhav1998aerospace}. 
LTL provides the mathematical rigor necessary to express critical safety invariants and liveness conditions~\cite{bauer2011runtime}, which are widely supported by existing model checkers and runtime verification tools.

Despite these advantages, the industrial adoption of formal methods is impeded by the formalization bottleneck: the manual translation of natural-language (NL) requirements into precise LTL properties. This process is labor-intensive and error-prone, requiring rare expertise in both aerospace control engineering and formal logic. Furthermore, recent research exploring Large Language Models (LLMs)~\cite{Minaee2024Survey,naveed2025comprehensive,wen2024automatically,su2024cfstra} for this task (\eg, NL2SPEC~\cite{Cosler2023nl2specIT}, NL2TL~\cite{Chen2023NL2TLTN}, NL2LTL~\cite{Fuggitti2023NL2LTLA}) has highlighted a fundamental limitation: while these tools succeed on synthetic benchmarks, they often fail on real industrial documents (as shown in our motivating example and experiments under the industrial setting).

The core challenge lies in the heterogeneous nature of industrial requirements. In practice, aerospace requirement documents contain highly specialized jargon, ambiguous abbreviations~\cite{bernstein1998design}, and implicit temporal and logical structures that are not explicitly stated in the text. Consequently, existing tools often produce incorrect or incomplete specifications, limiting their applicability in practice.

To address these challenges, we present \aero, a systematic engineering framework designed to automate the generation of LTL specifications from industrial aerospace requirements.
\aero leverages the reasoning capabilities of LLMs but enhances them through two key industrial innovations:
\begin{itemize}
    \vspace{-4.5pt}
    \item \spacekg: A domain-specific data dictionary that automatically ingests engineering artifacts (such as interface tables) to resolve technical jargon into precise, code-level atomic propositions.
    \vspace{2.5pt}
    \item \spacerdl: A structured template language that guides LLMs to uncover implicit temporal keywords and logical relations within the context of aerospace control logic.
    \vspace{-4.5pt}
\end{itemize}

Designed for seamless toolchain integration, \aero facilitates a direct connection with the subsequent verification tool to support batch processing and iterative feedback loops. 
The framework was evaluated using 79 production requirements from real-world spacecraft control systems. 
Experimental results demonstrate that \aero achieves 85\% precision and 88\% recall, substantially reducing manual formalization effort and establishing a practical, automated pathway for rigorous software assurance within the aerospace industry.

\section{Background and Motivation}\label{sec:Background}
\subsection{The complexity of Industrial Requirements}

\begin{table}[t]
\footnotesize
  \centering
    \setstretch{1.15}
    \setlength{\abovecaptionskip}{2.5pt}
    \setlength{\belowcaptionskip}{-2.5pt}
  \caption{Examples of Natural Language and Corresponding LTL Specifications}
  \label{tab:1}
  \resizebox{1\linewidth}{!}{%
    \begin{tabular}{ll}
      \toprule
      \rowcolor[rgb]{.851,.851,.851}
      \textbf{Natural Language} & \textbf{LTL Specification} \\
      \midrule
      \cite{Chen2023NL2TLTN} Once red, the light cannot become green next.
      & $G(\text{red} \rightarrow X \neg\text{green})$ \\

      \cite{Chen2023NL2TLTN} Once the light is red, it must remain red until it turns yellow.
      & $G(\text{red} \rightarrow \text{red} \, U \, \text{yellow})$ \\

      \cite{Cosler2023nl2specIT} If a holds then c is true until b.
      & $\text{a} \rightarrow (\text{c} \, U \, \text{b})$ \\

      \cite{Cosler2023nl2specIT} If b holds, next c holds until a holds or always c holds.
      & $\text{b} \rightarrow X ((\text{c} \, U \, \text{a}) \lor G \, \text{c})$ \\

      \cite{Fuggitti2023NL2LTLA} Navigate to the green room while avoiding landmark~1.
      & $(F \, \text{green}) \land G(\neg\text{landmark~1})$ \\

      \cite{Fuggitti2023NL2LTLA} Swing by landmark~1 before ending up in the red room.
      & $F(\text{landmark~1} \land (F \, \text{red}))$ \\
      \bottomrule
    \end{tabular}
  }
  \vspace{-5pt}
\end{table}

In academic research, the task of translating natural language to Linear Temporal Logic (LTL) is often simplified to mapping short, isolated sentences.
Table~\ref{tab:1} presents six representative examples adopted in prior work, including NL2LTL~\cite{Fuggitti2023NL2LTLA}, NL2SPEC~\cite{Cosler2023nl2specIT}, and NL2TL~\cite{Chen2023NL2TLTN}. 
In these examples, the correspondence between natural language statements and LTL specifications is explicit and largely unambiguous, rendering the translation task relatively straightforward.

\begin{figure*}[t]
\vspace{-2.5pt}
    \centering
    \begin{minipage}{1\textwidth}
        \centering
        \includegraphics[width=\linewidth]{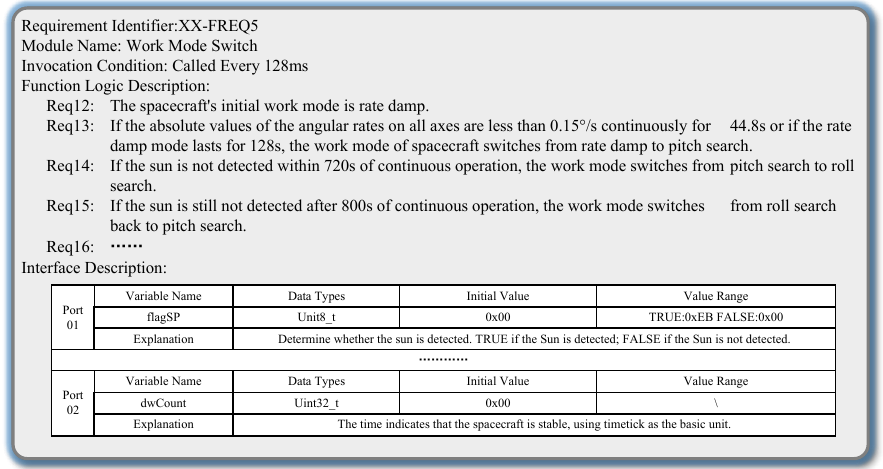}
        \subcaption{The requirement description for the Mode Switch Function Module of ACS-LEOS.\iffalse Clearly, the requirement descriptions natural language in real industrial scenarios are much more complex, with richer semantics and more diverse structures.\fi}
        \label{fig:figure1a}
    \end{minipage}\hfill
    \vspace{0.2cm}
    \begin{minipage}{1\textwidth}
        \centering
        \includegraphics[width=\linewidth]{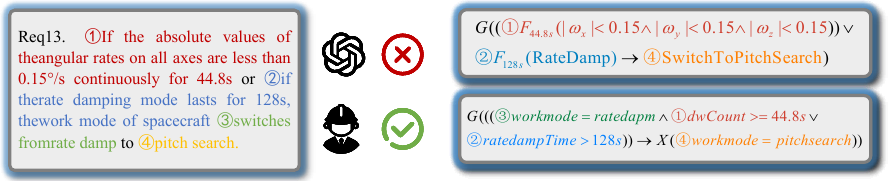}
        \subcaption{Misalignment of LLM-Generated LTL with Industrial Requirements. \iffalse The performance of LLMs in real industrial scenarios is not ideal, as the lack of necessary details results in LTL that does not align with the original intent of the requirement statements.\fi\iffalse The error symbol represents the specification generated by GPT-4, while the correct symbol indicates the manually revised specification.\fi}
        \label{fig:figure1b}
    \end{minipage}
    \setlength{\abovecaptionskip}{2.5pt}
    \setlength{\belowcaptionskip}{-12.5pt}
    \caption{The complexity and accuracy in NL requirement and its specification.}
    \label{fig:Figure1}
\end{figure*}

However, industrial requirements are fundamentally different, as illustrated by the example simplified from the ACS-LEOS\footnote{The Attitude Control Software for Low Earth Orbit Satellites (ACS-LEOS) is responsible for processing sensor data from sunlight and infrared instruments, managing transitions among multiple operating modes, and executing attitude adjustment tasks such as three-axis stabilization, spin stabilization, and sun-pointing control. \iffalse Its requirements document comprises 15\wc{7 module?} description units, each detailing the behavior and constraints of a specific functional module.\fi} in Fig.~\ref{fig:figure1a}.
They are not merely free-form prose but heterogeneous engineering artifacts characterized by three distinct layers of complexity:
(1) \textit{Multiple types of context}: A single functional requirement (\eg, \texttt{Req13}) cannot be parsed in isolation. Its meaning is bound to the \textit{Interface Description}, which specifies port-level information such as signal names, data types, initial values, and valid ranges, as well as the \textit{Invocation Condition}, which specifies the 128\,ms control cycle.
(2) Domain-Specific Semantics: Technical terminologies often act as placeholders for complex logic. In \texttt{Req13}, the phrase ``absolute values of the angular rates on all axes are less than 0.15°/s'' is an empirical condition for spacecraft stability. Without the domain knowledge found in the interface port explanations, an automated tool cannot map this intent to the correct variable, \texttt{dwCount}.
(3) Implicit Temporal Logic: Industrial requirements often omit explicit temporal operators, for example, in \texttt{Req13}, the term "switches from...to" implies a state transition that must occur at the next time step ($X$) once a sustaining condition is met.

\vspace{-15pt}

\subsection{Limitations of General-Purpose LLMs}\label{sec:limitation}

While LLMs like GPT-4 possess powerful reasoning capabilities, the generated LTL specifications are often unsatisfactory~\cite{cao2025informal,wen2024enchanting}.
As shown in Fig.~\ref{fig:figure1b}, although the model attempts to incorporate all relevant information in the requirement (\texttt{Req13}), the generated LTL specification deviates substantially from the intended semantics.
This mismatch stems primarily from three primary reasons. 
(1) \textit{Contextual Blindness to Engineering Artifacts.} LLMs struggle to connect natural-language requirements with their associated Interface Descriptions. 
This results in ``grounding errors,'' where the model cannot link a requirement to the actual system ports or data types defined in the document's tables.
(2) \textit{Insufficient Understanding of Domain-specific Terminology.} Industrial requirement statements frequently embed long technical expressions that domain experts interpret as single semantic units.
Lacking domain knowledge, LLMs tend to fragment such expressions into multiple isolated noun phrases, yielding atomic propositions that are semantically meaningless and difficult to maintain (\textit{e.g.}, {\color{red}\ding{172}} in \texttt{Req13}, shown in Fig.~\ref{fig:figure1b}).
(3) \textit{Incomplete Modeling of Implicit Temporal Constraints.} Industrial requirements often rely on implicit premises that are not explicitly marked by temporal keywords.
In \texttt{Req13}, the fragment {\color{green}\ding{174}}, ``\textit{switches from rate damp,}'' specifies a prerequisite mode that an LLM may fail to place on the left-hand side of a logical implication.
Furthermore, LLMs frequently omit essential temporal operators like $X$ (Next), which are necessary to indicate that a mode transition occurs in the subsequent control cycle. 
\textcolor{black}{It should be noted that, while the $X$ operator is often considered semantically unpredictable in general asynchronous systems, it is fully safe and deterministic in aerospace supervisory control software with fixed, strictly scheduled execution cycles, where $X$ precisely captures the state update in the immediate next cycle.}
These omissions lead to specifications that are syntactically valid yet semantically misaligned with the intended aerospace control logic.

\subsection{The Necessity of Domain-Integrated Formalization}

The limitations of general-purpose LLMs underscore a fundamental reality: in the aerospace domain, the ground truth of a requirement is not contained within the text alone, but is distributed across heterogeneous engineering artifacts. 
To bridge this gap, a framework must perform multi-source information synthesis, which motivates our two-pronged approach:
\begin{itemize}
    \vspace{-4.5pt}
    \item Semantic Grounding via \spacekg: To prevent the fragmentation of technical terms, the formalization process is grounded in a domain-specific data dictionary. By ingesting the Interface Description, \spacekg maps complex empirical constraints to precise, code-level atomic propositions. This ensures generated predicates are semantically consistent with actual system signals.
    \vspace{2.5pt}
    \item Structural Disambiguation via \spacerdl: To resolve the absence of explicit temporal keywords, \spacerdl provides structured semantic templates that regularize the LLM’s reasoning. These templates force implicit premises into an explicit intermediate format. This transforms ambiguous prose into a constrained logical structure that faithfully captures intended control laws.
    \vspace{-4.5pt}
\end{itemize}

By integrating \spacekg and \spacerdl, the formalization process shifts from a fragile "black-box" translation into a controllable engineering activity.


\section{The \aero Framework}\label{sec:framework}
\subsection{Overall Workflow}
\label{subsec:workflow}

\begin{figure*}[t]
\vspace{-2.5pt}
    \centering
    \includegraphics[width=1\textwidth]{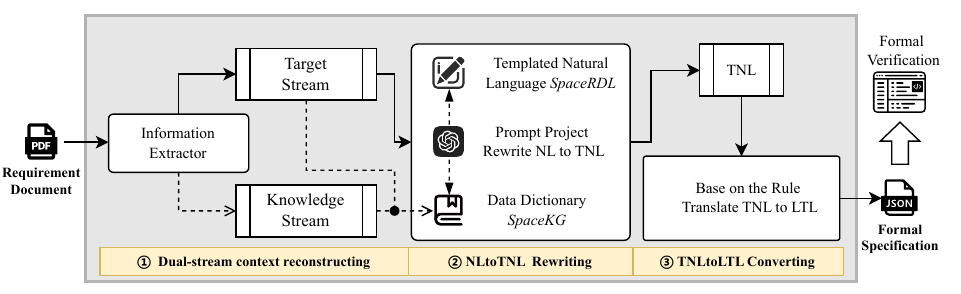}
    \setlength{\abovecaptionskip}{-7.5pt}
    \setlength{\belowcaptionskip}{-10pt}
    \caption{Workflow of \aero. \iffalse {\small Requirement statements are expressed in natural language. TNL represents templated natural language, and LTL represents linear temporal logic.}\fi}
    \label{fig:workflow1}
\end{figure*}

As illustrated in Fig.~\ref{fig:workflow1}, the \aero architecture operates in three consecutive stages, specifically addressing the challenges of \textit{contextual blindness to engineering artifacts}, \textit{insufficient understanding of domain-specific terminology}, \textit{incomplete modeling of implicit temporal constraints} identified in Section~\ref{sec:limitation}.

\smallskip
\textbf{Phase \ding{192}: Dual-stream Context Reconstructing.}
To address the issue of contextual blindness—where requirements are misinterpreted when analyzed as isolated sentences—\aero begins by processing the raw PDF document through a dual-stream extraction strategy. 
Using a layout-aware information extractor to parse complex industrial artifacts\footnote{We utilize the open-source \href{https://github.com/jsvine/pdfplumber}{\texttt{PDFplumber}} toolkit to robustly handle complex industrial layouts\iffalse, ensuring that every requirement statement is explicitly bound to its corresponding interface-level information\fi.}, the framework extracts two complementary information streams from the requirement document:
\begin{itemize}
\vspace{-5pt}
  \item \textit{Target Stream (Requirement Paragraphs):} 
  Instead of fragmented sentences, the framework extracts requirement statements as complete functional paragraphs. This preservation of paragraph structure allows LLMs to perceive the broader functional logic and explicit invocation conditions (\eg, control cycles). Providing this textual context is essential for LLMs to perform accurate subsequent temporal reasoning.

  \item \textit{Knowledge Stream (Interface Artifacts):} 
  Simultaneously, \aero harvests technical metadata from the \textit{Interface Description} tables associated with each module. 
  By extracting signal names, data types, and operational ranges, this stream provides the structured ``ground truth'' necessary for the subsequent construction of the \spacekg data dictionary. This ensures that linguistic expressions are grounded in actual engineering constraints.
  \vspace{-3.5pt}
\end{itemize}


\smallskip
\textbf{Phase \ding{193}: NL-to-TNL Rewriting.}
In the second stage, the framework transitions from open-ended natural language to \emph{Templated Natural Language} (TNL).
Here, LLMs' functions not as a stochastic generator, but as a semantic parser guided by two domain-specific components:
\begin{itemize}
\vspace{-4.5pt}
  \item \textit{Grounding via \spacekg}:
  To prevent terminology fragmentation, the previously extracted target and knowledge stream are ingested into \spacekg.
  Through a pipeline of BERT-based classification and expert-defined mapping, complex domain terms are normalized into precise, code-level \emph{Atomic Paradigms}.
\vspace{1.5pt}
  \item \textit{Structuring via \spacerdl}:
  To recover implicit temporal constraints, LLMs instantiate semantic structures defined by \spacerdl.
  By forcing requirements into predefined slots—such as \texttt{workmode}, \texttt{condition}, \texttt{timing}, \texttt{action} —the framework makes hidden premises and temporal operators (\eg, $X$ for next-cycle transitions) explicit within the TNL representation.
\vspace{-4.5pt}
\end{itemize}

\smallskip
\textbf{Phase \ding{194}: TNL-to-LTL Converting.}
The final stage eliminates inference stochasticity by replacing probabilistic LLMs generation with deterministic translation rules.
Since the TNL is already structurally consistent and semantically grounded, the framework applies fixed mapping logic to convert TNL constructs into syntactically valid LTL specifications.
This two-step process (NL $\rightarrow$ TNL $\rightarrow$ LTL) ensures that the final formal properties are syntactically valid and mathematically aligned with the intended aerospace control laws.

\subsection{\spacekg: Domain Grounding for Aerospace Requirements}
\label{subsec:spacekg}

\spacekg serves as the foundational semantic grounding layer of \aero, specifically designed to mitigate the terminology fragmentation of industrial documentation.
In aerospace requirements, complex engineering predicates—such as ``\textit{angular rates $< 0.15^\circ/s$}''—are frequently misinterpreted by text-only LLMs as isolated noun phrases. 
\spacekg resolves this by normalizing such expressions into unambiguous \textit{Atomic Paradigms} that link natural language intent directly to system signals.

\smallskip
\textbf{Knowledge Source and Construction.}
\spacekg is constructed from the \textit{Interface Description} sections identified during the context reconstruction phase. 
Unlike external ontologies, it derives knowledge from the same engineering artifacts that connect requirements to implementation, ensuring that the formalization is grounded in the exact variable definitions used in the software implementation.
The construction pipeline proceeds in three steps:
\begin{itemize}
\vspace{-5pt}
    \item \textit{Data Retrieval:} Harvesting variable names, data types, and explanatory semantics from interface tables.
    \item \textit{Intelligent Classification:} Utilizing a BERT-based terminology extractor to categorize entities into domain terms, variables, or concrete values.
    \item \textit{Expert Mapping:} Collaborating with domain experts to map complex empirical conditions (\eg, ``\textit{angular rates $< 0.15^\circ/s$}'') into standardized atomic expressions (\eg, \texttt{dwCount}).
\vspace{-5pt}
\end{itemize}

\begin{figure}[t]
\vspace{-2.5pt}
    \centering
    \includegraphics[width=0.95\textwidth]{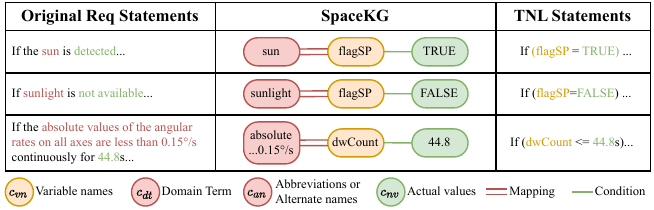}
    \setlength{\abovecaptionskip}{5pt}
    \setlength{\belowcaptionskip}{-10pt}
    \caption{Examples of requirement statements, corresponding \spacekg entries, and normalized atomic paradigms used in templated natural language (TNL).}
    \label{fig:spacekg}
\end{figure}

In practice, variable names are combined with their associated value constraints to form atomic paradigms suitable for formal reasoning.
For example, the empirical constraint ``\textit{the absolute values of the angular rates on all axes are less than $0.15^\circ$/s continuously for 44.8s}'' is normalized into the atomic expression ``$dwCount > 44.8s$'', which preserves the engineering intent while eliminating linguistic redundancy.
Figure~\ref{fig:spacekg} illustrates representative examples of original requirement statements, corresponding \spacekg entries, and the resulting normalized expressions.

\paragraph{\textbf{Formal Definition.}}
Formally, \spacekg is defined as a labeled graph \spacekg $(\bm{V}, \bm{E})$, where $\bm{V}$ is the set of nodes representing domain entities and $\bm{E} \subseteq \bm{V} \times \bm{V}$ is the set of edges representing semantic relationships among them.
Each node $v \in \bm{V}$ is mapped to a semantic class via a function $\tau: \bm{V} \rightarrow \bm{C}$.
The entity classes include domain terms ($c_{dt}$), abbreviations or alternate names ($c_{an}$), program variable names ($c_{vn}$), and concrete values ($c_{nv}$).
The values in $c_{nv}$ encompass not only booleans, integers, and floating-point numbers, but also predefined operational modes such as ``$RDSM$'', indicating the rate-damp work mode.

Edges in \spacekg are labeled by relationship types $r \in \bm{R}$ and mapped to pairs of entity classes via a function $\rho: \bm{R} \rightarrow \bm{C} \times \bm{C}$.
Two primary relationship types are defined.
\textcolor{black}{The first, $r_a$ (assignment/conditional relation), explicitly captures the logical or arithmetic constraints between program variables and their specific concrete values ; for example, mapping the informal condition ``\textit{sun is not detected}'' to the rigorous boolean evaluation ``\texttt{flagSP} $=$ \texttt{FALSE}''. 
The second, $r_m$ (semantic mapping relation), defines the semantic equivalences between formal program variables and various informal domain terms or aliases found in the text ; for instance, establishing that both natural language terms ``\textit{sun}'' and ``\textit{sunlight}'' semantically refer to the same underlying variable \texttt{flagSP}.}
The second, $r_m$, captures semantic mappings between program variables and domain terms or their aliases; for example, both ``\textit{sun}'' and ``\textit{sunlight}'' map to the variable \texttt{flagSP}.
Notably, in the aerospace domain, a domain term may correspond to a complete sentence rather than a single noun phrase, reflecting the complexity of engineering conventions in industrial requirement documents.

\subsection{\textit{SpaceRDL} and TNL: Explicitizing Implicit Temporal Semantics}
\label{subsec:spacerdl}

The second core contribution of \aero is \spacerdl, a structured requirement description language designed to make implicit temporal and logical assumptions explicit in aerospace control requirements.
As discussed in Section~\ref{sec:Background}, industrial aerospace requirements frequently rely on domain conventions—such as control cycles, physical constraints, and work-mode transitions—without explicitly stating temporal operators, which leads to ambiguity in automated formalization.

\spacerdl is derived from an empirical analysis of real aerospace control requirement documents.
Although requirement statements exhibit diverse surface forms, their underlying semantics consistently reflect the structured and mode-driven nature of spacecraft control systems.
This observation suggests that the apparent linguistic diversity of requirements can be systematically constrained by an explicit semantic structure aligned with the operational logic of aerospace control software.
Accordingly, \spacerdl provides a domain-driven semantic scaffold upon which heterogeneous natural-language requirements can be normalized and analyzed.

\smallskip
\textbf{\spacerdl Schema.}
Based on the above observation, \spacerdl defines a fixed set of semantic fields that explicitly capture the core elements of aerospace control logic:
\begin{equation*}
\vspace{-2.5pt}
\begin{aligned}
    \spacerdl = \{\texttt{workmode}, \texttt{condition}, \text{component}^*, 
    \texttt{SHALL}^*, \texttt{timing}^*, \texttt{action}^*\}
\vspace{-2.5pt}
\end{aligned}
\end{equation*}

Fields marked with an asterisk are mandatory.
The modal verb \texttt{SHALL*} is immutable and enforces declarative requirement expressions.
The \texttt{component*} field specifies the executor of the control action, while the optional \texttt{workmode} field constrains the operational scope under which the requirement applies; if omitted, the requirement is interpreted as globally applicable across all work modes.
The \texttt{timing*} field explicitly encodes temporal intent using a predefined set of keywords such as \textit{immediately}, \textit{next}, \textit{eventually}, \textit{always}, and \textit{until}, which correspond to standard LTL operators.
The \texttt{condition} field captures either trigger or sustaining semantics and supports logical composition through conjunctions and disjunctions.
\textcolor{black}{Finally, the \texttt{action*} field specifies the concrete operational behavior, state transition, or command execution that the target component is mandated to perform once the associated conditions and temporal constraints are satisfied.}
By enforcing these structural constraints, \spacerdl prevents implicit temporal premises from being silently omitted during formalization.

\begin{table*}[t]
  \centering
  \setstretch{1.2}
  \setlength{\abovecaptionskip}{2.5pt}
  \setlength{\belowcaptionskip}{-2.5pt}
  \caption{Representative SpaceRDL requirement templates and example TNL instantiations abstracted from aerospace industrial requirements.}
  \resizebox{\textwidth}{!}{
    \begin{tabular}{lll}
    \toprule
    \rowcolor[rgb]{ .851,  .851,  .851} \textbf{No.} & \textbf{Type} & \textbf{Requirement Templates and TNL Statements} \\
    \midrule
    \multirow{2}[2]{*}{1} & State & \textcolor{purple}{\textbf{Component}} shall always satisfy if ( \textcolor{purple}{\textbf{input\_state}} \& \textcolor{purple}{\textbf{condition}}) then next \textcolor{purple}{\textbf{output\_state}}. \\
          & Change   & 
          \textbf{`Communicator'} shall always satisfy if (\textbf{`signal\_lost'} \& \textbf{`mission\_phase'}) then next\textbf{`backup'}.
         \\
    \midrule
    \multirow{2}[2]{*}{2} & Bound & In \textcolor{purple}{\textbf{work\_mode}}, the \textcolor{purple}{\textbf{component}} shall always satisfy \textcolor{purple}{\textbf{bounds}}. \\
          & Check   & In \textbf{`pitch\_search mode'}, the \textbf{`gyroscope'} shall always satisfy \textbf{`roll\_target\_velocity = 0'}. \\
    \midrule
    \multirow{2}[2]{*}{3} & Work Mode & Upon ( \textcolor{purple}{\textbf{input\_work\_mode}} \& \textcolor{purple}{\textbf{condition}} ) \textcolor{purple}{\textbf{component}} shall at the next timepoint satisfy \textcolor{purple}{\textbf{output\_work\_mode}}. \\
          & Change   & Upon ( \textbf{`pitch\_search'} \& \textbf{`sun\_not\_found'} ) \textbf{`Spacecraft'} shall at the next timepoint satisfy \textbf{`roll\_search'}. \\
    
    \midrule
    \multirow{2}[2]{*}{4} & Command & Upon \textcolor{purple}{\textbf{command}} the \textcolor{purple}{\textbf{component}} shall \textcolor{purple}{\textbf{timing}} satisfy \textcolor{purple}{\textbf{action}}. \\
          & Process   & Upon \textbf{`command\_opcode = 29'} the software shall \textbf{`immediately'} satisfy \textbf{`retract\_landing\_gear\_command'}. \\

    \midrule
    \multirow{2}[2]{*}{5} & State & 
    
    In \textcolor{purple}{\textbf{work mode}} the \textcolor{purple}{\textbf{component}} shall immediately satisfy if \textcolor{purple}{\textbf{condition}} then \textcolor{purple}{\textbf{response}}.
    
    \\
          & Response   & In \textbf{roll\_hold mode} the \textbf{`Spacecraft'} shall immediately satisfy if \textbf{`roll\_angle\_up'} then \textbf{`roll\_hold'}. \\

    \midrule
    \multirow{2}[2]{*}{6} & State & The \textcolor{purple}{\textbf{component}} shall maintain \textcolor{purple}{\textbf{state}} unless \textcolor{purple}{\textbf{condition}}. \\
          & Maintenance   & The \textbf{`alarm system'} shall maintain \textbf{`active'} unless \textbf{`system\_disabled'}. \\

    \midrule
    \rowcolor[rgb]{ .851,  .851,  .851} \multirow{1}[1]{*}{7} &$\dots\dots\dots$  &$\dots\dots\dots$
    
    \\
    \bottomrule
    \end{tabular}}
  \label{tab:templates_example}%
  \vspace{-5pt}
\end{table*}

\smallskip
\textbf{Templated Natural Language (TNL)}.
Instantiating the fields of \spacerdl yields TNL, an intermediate representation that preserves human readability while making control semantics explicit and structured.
Rather than treating requirements as free-form text, TNL constrains each requirement to conform to a predefined semantic template derived from \spacerdl.

Empirical analysis further shows that the majority of aerospace industrial requirements can be categorized into a limited number of recurring semantic patterns.
These patterns correspond to fundamental control behaviors—such as state transitions, work-mode switches, boundary enforcement, command processing, and conditional responses—and form the basis of a compact template library.
Representative \spacerdl templates and example TNL instantiations are summarized in Table~\ref{tab:templates_example}.

By enforcing this template-based structure, TNL serves as a semantic contract between informal natural-language requirements and subsequent formal LTL specifications, substantially reducing ambiguity caused by implicit temporal assumptions and heterogeneous phrasing.

\section{Evaluation}\label{sec:eval}
The goal of our evaluation is to assess whether \aero can practically and reliably automate LTL specification generation from real industrial aerospace requirements.
Accordingly, our evaluation is organized around a real industrial case study, supported by quantitative analysis and targeted ablation experiments.

\subsection{Industrial Case and Dataset Construction}

Our evaluation is conducted on the \textit{Attitude Control Software for Low Earth Orbit Satellites} (ACS-LEOS), a production aerospace control system developed under strict safety and reliability constraints~\cite{ma2026integrating}.
ACS-LEOS is responsible for processing sensor data from sunlight and infrared instruments, managing transitions among multiple operating modes, and executing attitude adjustment tasks such as three-axis stabilization, spin stabilization, and sun-pointing control~\cite{zheng2026formal,ma2024towards}. 
We randomly selected nine control software engineering packages from the ACS-LEOS system as experimental subjects. Each package contains a complete set of requirement documents and the corresponding implementation code. 
Among them, the most recent package, the \emph{Sun Search Control Software} (SSCS), was chosen as the primary evaluation object. 
SSCS is a critical autonomous control program that enables a satellite to automatically search for and point toward the Sun when its attitude is unknown or abnormal, thereby restoring solar power supply and ensuring survivability. As a representative high-reliability autonomous control system, SSCS activates different control modes in a staged manner, with explicit temporal dependencies and state-triggering conditions between stages. These characteristics make temporal logic particularly suitable for formally specifying its requirements.

\smallskip
\textbf{Construction of \spacekg.}
The remaining eight historical engineering packages were used to construct the initial \spacekg. We first applied the \textit{Information Extractor} of \aero to the requirement documents of these packages, resulting in the extraction of 427 requirement statements and 635 domain-specific terms from interface description tables. A BERT-based terminology extractor was then employed to identify and classify domain entities using deep semantic features, including variables, system states, and operating modes. At this stage, the focus was on capturing stable aerospace concepts and commonly used abbreviations (\eg, mapping ``rate damping'' to \texttt{RDSM}).

In addition, aerospace engineers were involved to manually review and refine the automatically generated mappings between domain terms and atomic paradigms, particularly for interface variables with similar names but distinct semantics (\eg, \texttt{flagSP} versus \texttt{flagSPS}). The validated results were consolidated into the knowledge base for subsequent SSCS requirement analysis. The resulting initial version of \spacekg contains 407 predefined entity types.

Finally, to mitigate subtle terminology variations across different engineering documents, \spacekg incorporates an evolution mechanism. When processing new SSCS requirement documents, \aero activates an on-the-fly import module to handle project-specific primitives. In cases where the same domain term corresponds to different variable names, project-local artifacts are prioritized. Through this mechanism, \aero dynamically updates \spacekg, ensuring that complex empirical conditions (e.g., stability thresholds) are consistently mapped to the actual system ports defined in the current task document.

\smallskip
\textbf{Establishing Ground Truth.}
The SSCS requirement document spans 38 pages and contains 79 production requirements across seven functional modules.
To establish the ground-truth for evaluation, a panel of aerospace and formal methods experts manually formalized the 79 SSCS requirements. 
Of these, 76 were successfully translated into LTL, while three were excluded as high-level architectural summaries that do not contain temporal logic and cannot be translated into LTL.
Each expert independently performed the translation, followed by a cross-review and discussion phase to finalize the reference LTL specifications used for scoring.

\subsection{Evaluation Setup}

We implemented \aero as an automation-assisted toolchain that processes raw PDF requirement documents and produces LTL specifications enriched with structured metadata (\eg, module and component scope). 
The tool provides user interfaces for inspecting and correcting intermediate artifacts (extracted requirements, TNL, and generated LTL), supporting industrial and manual traceability.

\begin{table}[t]
\footnotesize
\setstretch{1.2}
\centering
\setlength{\abovecaptionskip}{2.5pt}
\setlength{\belowcaptionskip}{-2.5pt}
\caption{Effectiveness of \aero}

\resizebox{0.95\linewidth}{!}{%
\begin{tabular}{llcccccc}
\toprule
\rowcolor[rgb]{ .851,  .851,  .851}
\multicolumn{2}{l}{\textbf{Methods}} &
\textbf{Correct} &
\textbf{Missing} &
\textbf{Wrong} &
\textbf{Spurious} &
\textbf{Precision} &
\textbf{Recall} \\
\midrule

\multicolumn{2}{l}{\textbf{SimPro + DeepSeek}}  & 15 & 0 & 61 & 3 & 0.19 & 0.20 \\
\multicolumn{2}{l}{\textbf{SimPro + GPT-3.5}}   & 12 & 0 & 64 & 3 & 0.15 & 0.16 \\
\multicolumn{2}{l}{\textbf{SimPro + GPT-4o}}    & 27 & 0 & 49 & 1 & 0.35 & 0.36 \\

\multicolumn{2}{l}{\textbf{NL2LTL + DeepSeek}}  & 21 & 0 & 55 & 0 & 0.28 & 0.28 \\
\multicolumn{2}{l}{\textbf{NL2LTL + GPT-3.5}}   & 18 & 0 & 58 & 0 & 0.24 & 0.24 \\
\multicolumn{2}{l}{\textbf{NL2LTL + GPT-4o}}    & 37 & 0 & 39 & 0 & 0.49 & 0.49 \\

\multicolumn{2}{l}{\textbf{NL2SPEC + DeepSeek}} & 23 & 0 & 53 & 0 & 0.30 & 0.30 \\
\multicolumn{2}{l}{\textbf{NL2SPEC + GPT-3.5}}  & 23 & 0 & 53 & 0 & 0.30 & 0.30 \\
\multicolumn{2}{l}{\textbf{NL2SPEC + GPT-4o}}   & 46 & 0 & 30 & 0 & 0.61 & 0.61 \\
\midrule

\multicolumn{2}{l}{\textbf{\aero\,+\,DeepSeek}} & 64 & 0 & 12 & 3 & 0.81 & 0.84 \\
\multicolumn{2}{l}{\textbf{\aero\,+\,GPT-3.5}}  & 59 & 0 & 17 & 3 & 0.75 & 0.78 \\
\rowcolor[rgb]{ .851,  .851,  .851}
\multicolumn{2}{l}{\textbf{\aero\,+\,GPT-4o}} &
\textbf{67} & \textbf{0} & \textbf{9} & 3 & \textbf{0.85} & \textbf{0.88} \\
\bottomrule
\end{tabular}%
}

\label{tab:effect}

\begin{tablenotes}
\scriptsize
\item[*] SimPro means direct translation using LLMs with a zero-shot prompt.
\end{tablenotes}
\vspace{-7.5pt}
\end{table}

\smallskip
\textbf{Baselines.}
We compare \aero against several baseline methods across two dimensions: model backend and generation strategy.
For model backends, we choose three representative large language models: GPT-3.5, GPT-4o~\cite{openai2024gpt4o}, and {DeepSeek-V3}~\cite{zhang2024deepseek}.
For generation strategies, we consider three methods representing the current landscape of NL-to-LTL translation. (1) SimPro means direct translation using LLMs to assess the baseline "out-of-the-box" reasoning of LLMs; (2) {NL2LTL}~\cite{Fuggitti2023NL2LTLA}: a Python package developed by IBM Research that uses LLMs to translate NL instructions into LTL formulas. (3) {NL2SPEC}~\cite{Cosler2023nl2specIT} adopts a template-guided prompting approach that builds LTL expressions in stages, emphasizing interpretability and semantic traceability. 

Following common practice for fairness, we do not require baselines to adopt \spacekg naming conventions; if a baseline identifies the correct atomic proposition scope and temporal structure, the result is considered correct.

\smallskip
\textbf{Evaluation Metrics.}
We report \textit{precision} and \textit{recall} using statistical correctness criteria.
Each requirement is classified into one of four outcomes: \textit{Correct}, \textit{Wrong}, \textit{Missing}, or \textit{Spurious}.
A generated LTL property is counted as correct only if it matches the intended logical and temporal semantics; minor operator differences (\eg, $X$ vs.\ $F$) are considered \textit{wrong} due to their significant impact on verification outcomes.
If a requirement does not describe any checkable temporal behavior and therefore should not yield an LTL specification, then erroneously producing output is considered \textit{spurious}.
Conversely, if a requirement is expected to yield an LTL specification but the method fails to produce one, the outcome is classified as \textit{Missing}.

\subsection{Evaluation Result}

\smallskip
\textbf{Effectiveness.}
Table~\ref{tab:effect} summarizes end-to-end performance on the SSCS case study.
Overall, \aero consistently outperforms baseline methods under real industrial document complexity.
\aero + GPT-4o performs best with a precision of 85\% and a recall of 88\%. 
Across backends, \aero attains 85\%, 75\%, and 81\% precision (GPT-4o, GPT-3.5, DeepSeek-V3 respectively), showing that the framework generalizes beyond a single model while benefiting from stronger LLM reasoning.

In contrast, baseline methods degrade substantially under industrial document complexity. 
A key reason is that industrial aerospace requirements cannot be interpreted as standalone sentences: correct formalization requires grounding technical jargon and numeric constraints to interface variables, and recovering temporal/logical structure that is often implicit. 
SimPro and NL2SPEC struggle primarily because they treat requirements as isolated text, failing to resolve the technical jargon and implicit dependencies defined in the document's interface description. 
While NL2LTL and NL2SPEC avoid \textit{Spurious} outputs by being more conservative, \aero’s proactive two-stage rewriting process (NL$\rightarrow$TNL$\rightarrow$LTL) exposes implicit conditions and timing intent in TNL, and then applies deterministic translation rules. 
This design improves coverage while preserving precision, with only a small increase in spurious outputs on highly generalized, non-checkable statements.

\smallskip
\textbf{Ablation Study.}
We conducted ablation studies to quantify the necessity of \spacekg and \spacerdl in the formalization pipeline and to analyze the underlying causes of performance degradation when each component is removed, as shown in Table~\ref{tab:ablation}. 

\begin{table*}[t]
\footnotesize
\setstretch{1.15}
    \centering
    \setlength\tabcolsep{5pt}
    \setlength{\abovecaptionskip}{2.5pt}
    \setlength{\belowcaptionskip}{-2.5pt}
  \caption{Ablation Study of \spacekg and \spacerdl Using GPT-4o.}
    \begin{tabular}{llcccccc}
    \toprule
    \rowcolor[rgb]{ .851,  .851,  .851}\multicolumn{2}{l}{\textbf{Components}} & \textbf{Correct} & \textbf{Missing} & \textbf{Wrong} & \textbf{Spurious} & \textbf{Precision} & \textbf{Recall} \\
    \midrule
    \multicolumn{2}{l}{\aero} & 67 & 0 & 9 & 3     & 0.85 & 0.88 \\
    \multicolumn{2}{l}{w/o \spacekg} & 53    & 0 & 23     & 1     & 0.69  & 0.70 \\
    \multicolumn{2}{l}{w/o \spacerdl} & 41    & 0 & 35     & 3     & 0.52  & 0.54 \\
    \bottomrule
    \end{tabular}%
  \label{tab:ablation}%
  \vspace{-5pt}
\end{table*}%

\begin{figure}[t]
    \centering
    \begin{minipage}[t]{0.49\textwidth}
        \centering
        \includegraphics[height=0.22\textheight]{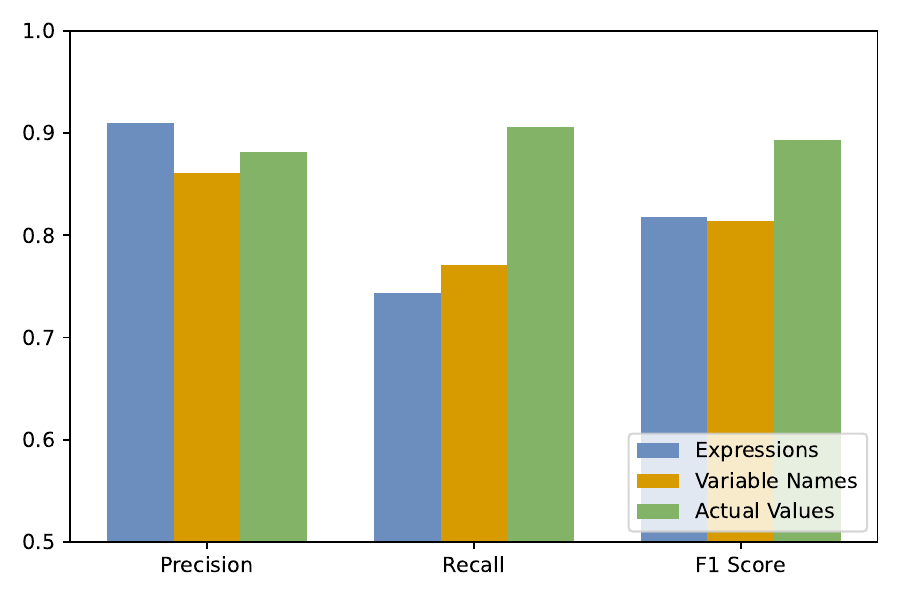}
        \caption{Accuracy of \spacekg.}
        \label{fig:spacekg_perf}
    \end{minipage}
    \begin{minipage}[t]{0.49\textwidth}
        \centering
        \includegraphics[height=0.22\textheight]{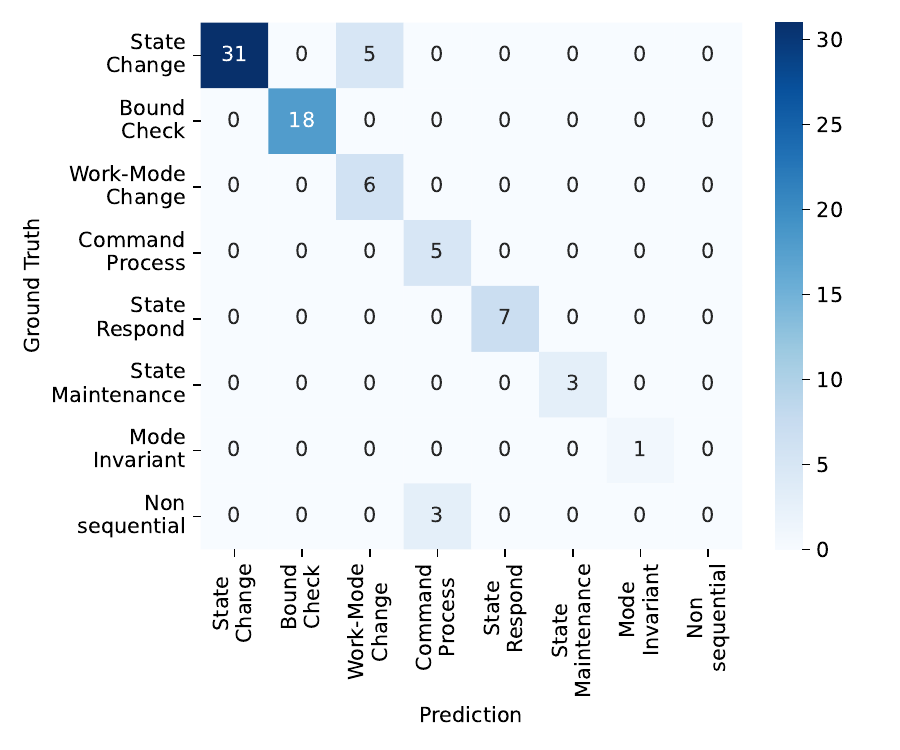}
        \caption{Accuracy of \spacerdl.}
        \label{fig:SpaceRDL_classify}
    \end{minipage}
 \vspace{-10pt}
\end{figure}

As shown in the results, removing \spacekg causes precision and recall to drop to 69\% and 70\%, respectively. The primary errors are grounding failures, where domain-specific terms are either fragmented into semantically meaningless propositions or mapped to incorrect yet plausible variables, such as confusing \texttt{flagSP} with \texttt{flagSPS}. To better understand these failures, we further analyze the extraction accuracy of \spacekg on the 79 SSCS requirements in Fig.~\ref{fig:spacekg_perf}. The extracted domain terms, variables, and expressions all demonstrate F1 scores exceeding 0.8, proving \spacekg's critical role in preventing grounding errors through effective terminology identification and normalization.

Subsequently, removing \spacerdl leads to a more severe performance degradation, with precision and recall dropping to 52\% and 54\%, respectively. These errors mostly stem from incorrect temporal structuring, including missing prerequisites on the left-hand side of implications and the incorrect placement or omission of temporal operators such as $X$. A confusion matrix further analyzes the model's ability to select appropriate semantic templates for requirement statements, as shown in Fig.~\ref{fig:SpaceRDL_classify}. While \aero correctly classifies 71 statements, several \textit{State-Transition} requirements are misclassified as \textit{Work-Mode-Change}, and non-checkable requirements tend to be forcibly assigned a template. These misclassifications directly account for the temporal-structure errors observed in the ablation results. These findings confirm that both \spacekg and \spacerdl are essential to the accuracy and reliability of \aero.

\subsection{End-to-end Case Study and Toolchain Integration}
\label{subsec:case_study}

To demonstrate practical usability, we present a detailed walkthrough using \texttt{Req39} from the SSCS \textit{Mode Switch} module. 

\vspace{5pt}
(\texttt{Req39}).
\textit{In the event that the BeiDou signal is unavailable for a duration exceeding 12.8s while operating in inertial navigation mode, the navigation computer is required to initiate an immediate transition to the star tracker system.}

\smallskip
\textbf{Step 1: Context Reconstruction and Knowledge Grounding.}
The workflow begins by jointly reconstructing textual and engineering context from the SSCS requirement document.
The \textit{Information Extractor} identifies the requirement paragraph of \texttt{Req39} as the \textit{Target Stream}, while simultaneously extracting the associated \textit{Interface Description} table as the \textit{Knowledge Stream}. 
Based on this information, \aero determines that the BeiDou signal availability is represented by the boolean variable \texttt{flagBD}, the duration timer corresponds to the floating-point variable \texttt{deTCount}, and the star tracker system status is denoted by \texttt{flagSTS}. 
Based on this project-local evidence, \aero performs a just-in-time \textbf{delta refresh} of \spacekg, augmenting its domain-level knowledge (\eg, mapping ``inertial navigation mode'' to \texttt{subMode = INM}) with SSCS-specific signal grounding.
This step is essential for industrial correctness. 
Without ingesting interface definitions, a text-only approach cannot associate phrases such as ``duration exceeding 12.8s'' with the specific runtime variable \texttt{deTCount}, resulting in an underspecified or unverifiable property.

\smallskip
\textbf{Step 2: Semantic Rewriting (NL $\rightarrow$ TNL).}
With the project-adapted \spacekg in place, \aero invokes the LLM-based semantic rewriter to transform the original natural-language requirement into a structured intermediate representation.
During this process, domain phrases are deterministically grounded: ``BeiDou signal is unavailable'' is mapped to \texttt{flagBD = FALSE}, and ``duration exceeding 12.8s'' is resolved as the numeric constraint \texttt{deTCount > 12.8}.
Meanwhile, guided by \spacerdl templates, the rewriter identifies the operational context (``inertial navigation mode'') and the implicit timing keyword ``immediate''.
These elements trigger a \textit{State--Response} template instantiation, yielding the following templated natural language specification:
\begin{equation*}
\vspace{-2.5pt}
\begin{aligned}
\texttt{TNL39: } 
\textit{In (}\texttt{subMode} \textit{=} \texttt{INM}\textit{), the navigation computer shall immediately satisfy} \\
\textit{if (}\texttt{flagBD} \textit{=} \texttt{FALSE} \;\textit{$\&$} \;\texttt{deTCount > 12.8}\textit{)} \\
\textit{then (}\texttt{flagSTS} \textit{=} \texttt{TRUE}\textit{).}
\vspace{-2.5pt}
\end{aligned}
\end{equation*}

\smallskip
\textbf{Step 3: Deterministic Formal Synthesis (TNL $\rightarrow$ LTL).}
\textcolor{black}{In the third stage, the instantiated semantic structure (\texttt{TNL39}) is passed to the deterministic synthesis engine.
To ensure rigorous formalization, \aero avoids probabilistic NL-to-LTL generation; instead, it applies formally defined \spacerdl translation rules that map each \texttt{TNL} field to its corresponding LTL construct , thereby eliminating semantic ambiguity.}
In particular, the keyword ``immediately'' is interpreted as a concurrent state implication rather than a future eventuality, and thus no $F$ (Eventually) operator is introduced.
The resulting formal specification is:
\begin{equation*}
\vspace{-4.5pt}
\begin{aligned}
\texttt{LTL39: } 
G(& (\texttt{subMode} = \texttt{INM} \land \texttt{flagBD} = \texttt{FALSE} \land \texttt{deTCount > 12.8}) \\
      & \rightarrow (\texttt{flagSTS = TRUE}) )
\end{aligned}
\vspace{-4.5pt}
\end{equation*}

\smallskip
For comparison, a representative text-only baseline such as NL2SPEC abstracts domain predicates into uninterpreted symbols and models the response using a future operator.
For instance, the original condition “the BeiDou signal is unavailable for a duration exceeding 12.8s” is reduced to a symbolic atom (e.g., $a$), while the response “initiate an immediate transition” is formalized as an eventuality, yielding a specification of the form $G((a \land b) \rightarrow F\,c)$.
This formulation omits the concrete timing threshold ($>12.8$s) and weakens the immediacy constraint by deferring the response to an unspecified future point, rendering the resulting property unsuitable for direct verification in a control-critical setting.


\smallskip
\textbf{Step 4: Direct Toolchain Integration and Verification.}
The final step evaluates whether the generated LTL specifications can be directly consumed by an existing industrial verification toolchain.
The synthesized property \texttt{LTL39}, together with other correctly generated specifications, is used as drop-in input to the TRACE~\cite{TRACE} framework for trace-based runtime verification\footnote{TRACE, developed by the Eindhoven University of Technology in the Netherlands, is an automated verification tool that takes LTL formulas as input and utilizes trace-based runtime verification technology to analyze system behavior by checking the execution traces of the system.}.
TRACE takes LTL formulas as input and performs symbolic execution using KLEE~\cite{cadar2008klee,cadar2021klee} to generate execution paths, while GDB~\cite{van2019program} is employed to collect concrete execution traces, which are subsequently checked against the temporal properties.

\begin{table*}[t]
\footnotesize
\setstretch{1.15}
    \centering
    \setlength\tabcolsep{20pt}
    \setlength{\abovecaptionskip}{2.5pt}
    \setlength{\belowcaptionskip}{-2.5pt}
  \caption{Verification results of Generated LTL specifications using TRACE.}
    \begin{tabular}{l|c}
    \toprule
    \rowcolor[rgb]{ .851,  .851,  .851} \multicolumn{2}{c}{\textbf{Sun Search Control System}} \\
    \midrule
    \textbf{Module Name} & \textbf{LTL properties} \\
    \midrule
    \texttt{Remote Control Command Processing} & 17/17 \\
    \texttt{Gyroscope Data Processing} & 13/15 \\
    \texttt{Attitude Determination} & 10/12 \\
    \texttt{Control Calculation} & 5/5 \\
    \texttt{Mode Switch} & 6/6 \\
    \texttt{Fault Handling} & 8/8 \\
    \texttt{Thruster Output} & 4/4 \\
    \midrule
    \rowcolor[rgb]{ .851,  .851,  .851} \textbf{Total} & 63/67 \\
    \bottomrule
    \end{tabular}    
  \label{tab:rq3}%
  \vspace{-5pt}
\end{table*}%

\smallskip
\textbf{SSCS Verification Results.}
In the SSCS evaluation, \aero is applied to 76 industrial requirements without manual intervention, of which 67 are correctly translated into LTL.
As summarized in Table~\ref{tab:rq3},  the downstream TRACE~\cite{TRACE} toolchain successfully verifies 63 of these 67 properties across diverse functional modules.
These results demonstrate that \aero produces syntactically valid and semantically precise specifications that are immediately usable by existing verification tools, effectively completing the end-to-end workflow from natural-language requirements to executable software verification.

\subsection{Discussion on Scalability and Performance}

\textcolor{black}{While the current evaluation demonstrates the effectiveness of \aero on the 79 production requirements of the SSCS module, scaling this approach to the full set of requirements typically found in safety-critical aerospace platforms presents a distinct engineering challenge in and of itself. Complete systems often comprise over 1,000 behavioral requirements, alongside correspondingly massive and complex interface description documents.}

\textcolor{black}{In such large-scale industrial scenarios, the performance and scalability of the automated toolchain become critical considerations. The primary computational overheads in \aero stem from the dynamic construction of \spacekg and the LLM inference latency. To effectively process thousands of requirements, the layout-aware information extraction and \spacekg entity normalization must be highly optimized. This could be achieved through incremental knowledge base updates—selectively parsing modified interface tables rather than performing full-document re-ingestion. Furthermore, while \aero currently supports batch processing for downstream verification tools like TRACE, the LLM-based NL-to-TNL rewriting phase will necessitate advanced parallel processing and modularized semantic parsing. Addressing these performance bottlenecks and ensuring that generation times remain practical for continuous integration and iterative engineering feedback loops constitute an important direction for scaling \aero to broader, system-level aerospace projects.}

\section{Related Work}\label{sec:related}
Recent works explore translating natural language requirements into Linear Temporal Logic using Large Language Models~\cite{English2025GraFT,Cosler2023nl2specIT,Chen2023NL2TLTN,Fuggitti2023NL2LTLA,liu2022lang2ltl,xu2024learning}. While effective on carefully constructed benchmarks, these approaches often fail in industrial settings such as robotics~\cite{Liu2023Lang2LTL,Liu2024Lang2LTL2GS}, multi-agent systems~\cite{Xu2024Nl2Hltl2Plan}, and software verification~\cite{Olausson2023DemystifyingGS_ref34}. Our analysis of these industrial failures reveals three recurring patterns:

\textit{1. Contextual blindness to engineering artifacts}. Industrial specifications distribute state-transition logic across long documents. In telecommunications standards such as 3GPP~\cite{Zhang2025SpecGPT}, related conditions may appear in distant sections or appendices, making global consistency difficult for LLMs to maintain. As a result, extracted state machines are often fragmented. This challenge underscores the fundamental principle of the "Triptych" approach proposed by Dines Bjørner~\cite{Bjorner2006SE3}, which asserts that a rigorous domain description is a mandatory prerequisite for any requirements prescription~\cite{Bjorner2016From}. Without an explicit domain model to provide the necessary context, requirements become isolated and prone to misinterpretation. Wu et al.~\cite{Wu2026Unlocking} address this problem through an iterative, business-logic-driven framework that reconstructs implicit requirements by repeatedly aligning partial specifications with system-level objectives.

\textit{2. Insufficient understanding of domain-specific terminology}~\cite{Shih2025FLAG,Meloche2023Symboleo,Ma2025OnionL}. In legal texts, LLMs frequently confuse obligation and permission, mistranslating \textit{shall} or ignoring \textit{unless} clauses~\cite{Meloche2023Symboleo}. Similar issues arise in hardware and protocol specifications, where global state-machine semantics in standards such as Arm are often misinterpreted~\cite{Shih2025FLAG,Zhang2025SpecGPT}. Such terminological ambiguity often stems from the lack of a formal domain ontology, a gap that Bjørner suggests filling with domain engineering to capture the intrinsic properties of the application environment~\cite{Bjorner2000Domain}. Surveys further report frequent errors in temporal semantics~\cite{Wu2024Survey,Abdollahi2024HardwareDA}, such as treating \textit{stable} as a static predicate rather than a clocked condition in SystemVerilog Assertions. Huang et al.~\cite{Huang2026Modeling} mitigate these issues via layered semantic analysis that incrementally extracts atomic propositions, mode constraints, and timing information.

\textit{3. Incomplete modeling of implicit temporal constraints}~\cite{Tang2025VLTLBench}. Sequence-to-sequence models struggle with nested conditions, recursive triggers, and mode-dependent logic~\cite{English2025GraFT}. Flat translations fail to capture hierarchical temporal structure, and text-based metrics such as BLEU are insufficient, as small operator changes (\eg, $F$ vs. $G$) can drastically alter semantics~\cite{Tang2025VLTLBench,English2025GraFT}. Trajectory-based benchmarks like VLTL-Bench reveal poor semantic equivalence despite high textual similarity~\cite{Tang2025VLTLBench}. Consequently, recent work emphasizes structured and verifiable methods, including logic-aware decoding~\cite{English2025GraFT}, interactive guidance~\cite{Cosler2023nl2specIT,mendoza2024synttl}, conformal guarantees~\cite{Wang2025ConformalNL2LTL}, and intermediate semantic representations~\cite{Huang2026Modeling,Wu2026Unlocking}.

\section{Conclusions}\label{sec:conclu}
This paper presents \aero, an automated framework for extracting LTL specifications from aerospace requirement documents. By utilizing a large language model along with domain-specific knowledge and standardized templates, the framework improves the accuracy of temporal relationship extraction. \aero integrates seamlessly with existing verification tools like TRACE and CPAchecker and is already being applied in real aerospace software projects. Experimental results and case studies demonstrate its effectiveness, while ablation experiments highlight the challenges in rewriting natural language into templated forms.



\textcolor{black}{Future work will focus on three directions to further improve the framework’s generalizability and reliability. First, since \spacekg is constructed from common engineering artifacts (e.g., interface description tables) rather than domain-specific hardcoding, the toolchain is inherently extensible beyond aerospace. We will apply it to other safety-critical domains with similar documentation standards, such as automotive and medical systems. Second, beyond control-related state transitions and work-mode logic, we will expand \aero to support a wider range of specifications, including functional, safety, and security requirements. Finally, we will further enrich and formalize the \spacekg model by systematically encoding domain invariants into its core structure, thereby strengthening the rigor and validity of the automated verification process.}

\section*{Acknowledgements}
This work was supported in part by the National Natural Science Foundation of China (Nos. 62192730, 62192734, 62192735, 62302375, 62472339), the China Postdoctoral Science Foundation funded project (No. 2023M723736), the Basic Research Foundation of Shenzhen City (No. JCYJ20250604184202003), and the CCF-Huawei Populus Grove Fund (No. CCF-HuaweiFM202507).


\newpage
\bibliographystyle{unsrt} 
\bibliography{Tex/reference}

\begin{thebibliography}{10}

\bibitem{rozier2011linear}
Kristin~Y Rozier.
\newblock Linear temporal logic symbolic model checking.
\newblock {\em Computer Science Review}, 5(2):163--203, 2011.

\bibitem{merhav1998aerospace}
Shmuel Merhav.
\newblock {\em Aerospace sensor systems and applications}.
\newblock Springer Science \& Business Media, 1998.

\bibitem{bauer2011runtime}
Andreas Bauer, Martin Leucker, and Christian Schallhart.
\newblock Runtime verification for ltl and tltl.
\newblock {\em ACM Transactions on Software Engineering and Methodology (TOSEM)}, 20(4):1--64, 2011.

\bibitem{Minaee2024Survey}
Shervin Minaee, Tomas Mikolov, Narjes Nikzad, Meysam Chenaghlu, Richard Socher, Xavier Amatriain, and Jianfeng Gao.
\newblock Large language models: A survey.
\newblock {\em arXiv preprint arXiv:2402.06196}, 2024.

\bibitem{naveed2025comprehensive}
Humza Naveed, Asad~Ullah Khan, Shi Qiu, Muhammad Saqib, Saeed Anwar, Muhammad Usman, Naveed Akhtar, Nick Barnes, and Ajmal Mian.
\newblock A comprehensive overview of large language models.
\newblock {\em ACM Transactions on Intelligent Systems and Technology}, 16(5):1--72, 2025.

\bibitem{wen2024automatically}
Cheng Wen, Yuandao Cai, Bin Zhang, Jie Su, Zhiwu Xu, Dugang Liu, Shengchao Qin, Zhong Ming, and Tian Cong.
\newblock Automatically inspecting thousands of static bug warnings with large language model: How far are we?
\newblock {\em ACM Transactions on Knowledge Discovery from Data}, 18(7):1--34, 2024.

\bibitem{su2024cfstra}
Jie Su, Liansai Deng, Cheng Wen, Shengchao Qin, and Cong Tian.
\newblock Cfstra: Enhancing configurable program analysis through llm-driven strategy selection based on code features.
\newblock In {\em International Symposium on Theoretical Aspects of Software Engineering}, pages 374--391. Springer, 2024.

\bibitem{Cosler2023nl2specIT}
Matthias Cosler, Christopher Hahn, Daniel Mendoza, Frederik Schmitt, and Caroline Trippel.
\newblock nl2spec: Interactively translating unstructured natural language to temporal logics with large language models.
\newblock In {\em Proceedings of the 35th International Conference on Computer Aided Verification}, volume 13965 of {\em Lecture Notes in Computer Science}, pages 383--396, Paris, France, 2023. Springer.

\bibitem{Chen2023NL2TLTN}
Yongchao Chen, Rujul Gandhi, Yang Zhang, and Chuchu Fan.
\newblock Nl2tl: Transforming natural languages to temporal logics using large language models.
\newblock In {\em Proceedings of the Conference on Empirical Methods in Natural Language Processing}, pages 15880--15903, Singapore, 2023. Association for Computational Linguistics.

\bibitem{Fuggitti2023NL2LTLA}
Francesco Fuggitti and Tathagata Chakraborti.
\newblock Nl2ltl - a python package for converting natural language (nl) instructions to linear temporal logic (ltl) formulas.
\newblock In {\em Proceedings of the 37th {AAAI} Conference on Artificial Intelligence, the 35th Conference on Innovative Applications of Artificial Intelligence, the 13th Symposium on Educational Advances in Artificial Intelligence}, pages 16428--16430, Washington, DC, USA, 2023. {AAAI} Press.

\bibitem{bernstein1998design}
Joshua~Ian Bernstein.
\newblock {\em Design methods in the aerospace industry: looking for evidence of set-based practices}.
\newblock PhD thesis, Massachusetts Institute of Technology, 1998.

\bibitem{cao2025informal}
Jialun Cao, Yaojie Lu, Meiziniu Li, Haoyang Ma, Haokun Li, Mengda He, Cheng Wen, Le~Sun, Hongyu Zhang, Shengchao Qin, et~al.
\newblock From informal to formal--incorporating and evaluating llms on natural language requirements to verifiable formal proofs.
\newblock In {\em Proceedings of the 63rd Annual Meeting of the Association for Computational Linguistics (Volume 1: Long Papers)}, pages 26984--27003, 2025.

\bibitem{wen2024enchanting}
Cheng Wen, Jialun Cao, Jie Su, Zhiwu Xu, Shengchao Qin, Mengda He, Haokun Li, Shing-Chi Cheung, and Cong Tian.
\newblock Enchanting program specification synthesis by large language models using static analysis and program verification.
\newblock In {\em International Conference on Computer Aided Verification}, pages 302--328. Springer, 2024.

\bibitem{ma2026integrating}
Zhi Ma, Cheng Wen, Bin Yu, and Jie Su.
\newblock Integrating ensemble learning and large language models for efficient formal verification of ip-based aerospace systems.
\newblock {\em Information Fusion}, 125:103466, 2026.

\bibitem{zheng2026formal}
Shaolin Zhang, HongJin Liu, Zhi Ma, Xiao Liang, and Cheng Wen.
\newblock Formal verification of aerospace software ip components: A multi-tool case study.
\newblock In {\em Proceedings of the 2026 6th International Conference on Computer Network Security and Software Engineering}, pages 1--9, 2026.

\bibitem{ma2024towards}
Zhi Ma, Cheng Wen, Jie Su, Ming Zhao, Bin Yu, Xu~Lu, and Cong Tian.
\newblock Towards practical requirement analysis and verification: A case study on software ip components in aerospace embedded systems.
\newblock {\em arXiv preprint arXiv:2404.00795}, 2024.

\bibitem{openai2024gpt4o}
OpenAI.
\newblock Gpt-4o technical report.
\newblock \url{https://openai.com/index/gpt-4o}, 2024.
\newblock Accessed: May 2025.

\bibitem{zhang2024deepseek}
Shuai Zhang, Haisen Zhao, et~al.
\newblock Deepseek-vl: Scaling vision-language models with vision token learner.
\newblock {\em arXiv preprint arXiv:2405.07927}, 2024.

\bibitem{TRACE}
Martijn Hendriks, Marc Geilen, Amir Reza~Baghban Behrouzian, Twan Basten, Hadi~Alizadeh Ara, and Dip Goswami.
\newblock Checking metric temporal logic with trace.
\newblock {\em 2016 16th International Conference on Application of Concurrency to System Design (ACSD)}, pages 19--24, 2016.

\bibitem{cadar2008klee}
Cristian Cadar, Daniel Dunbar, Dawson~R Engler, et~al.
\newblock Klee: Unassisted and automatic generation of high-coverage tests for complex systems programs.
\newblock In {\em OSDI}, volume~8, pages 209--224, 2008.

\bibitem{cadar2021klee}
Cristian Cadar and Martin Nowack.
\newblock Klee symbolic execution engine in 2019.
\newblock {\em International Journal on Software Tools for Technology Transfer}, 23:867--870, 2021.

\bibitem{van2019program}
Jo~Van~Hoey.
\newblock Program analysis with a debugger: Gdb.
\newblock In {\em Beginning x64 Assembly Programming: From Novice to AVX Professional}, pages 21--33. Springer, 2019.

\bibitem{English2025GraFT}
William English, Dominic Simon, Sumit~Kumar Jha, and Rickard Ewetz.
\newblock Grammar-forced translation of natural language to temporal logic using llms.
\newblock In {\em International Conference on Machine Learning}, 2025.

\bibitem{liu2022lang2ltl}
Jason~Xinyu Liu, Ziyi Yang, Benjamin Schornstein, Sam Liang, Ifrah Idrees, Stefanie Tellex, and Ankit Shah.
\newblock Lang2ltl: Translating natural language commands to temporal specification with large language models.
\newblock In {\em Workshop on Language and Robotics at CoRL 2022}, 2022.

\bibitem{xu2024learning}
Yilongfei Xu, Jincao Feng, and Weikai Miao.
\newblock Learning from failures: Translation of natural language requirements into linear temporal logic with large language models.
\newblock In {\em 2024 IEEE 24th International Conference on Software Quality, Reliability and Security (QRS)}, pages 204--215. IEEE, 2024.

\bibitem{Liu2023Lang2LTL}
Jason~Xinyu Liu, Ziyi Yang, Benjamin Schornstein, Sam Liang, Ifrah Idrees, Stefanie Tellex, and Ankit Shah.
\newblock Lang2ltl: Translating natural language commands to temporal specification with large language models.
\newblock In {\em Workshop on Language and Robotics at CoRL 2022}, 2022.

\bibitem{Liu2024Lang2LTL2GS}
Jason~Xinyu Liu, Ankit Shah, George~Dimitri Konidaris, Stefanie Tellex, and David Paulius.
\newblock Lang2ltl-2: Grounding spatiotemporal navigation commands using large language and vision-language models.
\newblock {\em 2024 IEEE/RSJ International Conference on Intelligent Robots and Systems (IROS)}, pages 2325--2332, 2024.

\bibitem{Xu2024Nl2Hltl2Plan}
Shaojun Xu, Xusheng Luo, Yutong Huang, Letian Leng, Ruixuan Liu, and Changliu Liu.
\newblock Nl2hltl2plan: Scaling up natural language understanding for multi-robots through hierarchical temporal logic task specifications.
\newblock {\em IEEE Robotics and Automation Letters}, 10:10482--10489, 2024.

\bibitem{Olausson2023DemystifyingGS_ref34}
Theo~X. Olausson, Jeevana~Priya Inala, Chenglong Wang, Jianfeng Gao, and Armando Solar-Lezama.
\newblock Demystifying gpt self-repair for code generation.
\newblock {\em ArXiv}, abs/2306.09896, 2023.

\bibitem{Zhang2025SpecGPT}
Miao Zhang, Runhan Feng, Hongbo Tang, Yu~Zhao, Jie Yang, Hang Qiu, and Qi~Liu.
\newblock Automated extraction of protocol state machines from 3gpp specifications with domain-informed prompts and llm ensembles.
\newblock {\em ArXiv}, abs/2510.14348, 2025.

\bibitem{Bjorner2006SE3}
Dines Bj{\o}ner.
\newblock {\em Software Engineering 3: Domains, requirements, and software design}.
\newblock Springer, 2006.

\bibitem{Bjorner2016From}
Dines Bj{\o}rner.
\newblock From domain descriptions to requirements prescriptions.

\bibitem{Wu2026Unlocking}
Zhujun Wu, Xiaohong Chen, Zhi Jin, Ming Hu, and Dongming Jin.
\newblock Unlocking the silent needs: Business-logic-driven iterative requirements auto-completion.
\newblock In {\em 2026 IEEE/ACM 48th International Conference on Software Engineering (ICSE '26)}, page~12, 2026.

\bibitem{Shih2025FLAG}
Yu-An Shih, Annie Lin, Aarti Gupta, and Sharad Malik.
\newblock Flag: Formal and llm-assisted sva generation for formal specifications of on-chip communication protocols.
\newblock {\em ArXiv}, abs/2504.17226, 2025.

\bibitem{Meloche2023Symboleo}
Regan Meloche, Daniel Amyot, and John Mylopoulos.
\newblock Towards legal contract formalization with controlled natural language templates.
\newblock In {\em 2023 IEEE 31st International Requirements Engineering Conference (RE)}, pages 317--322. IEEE, 2023.

\bibitem{Ma2025OnionL}
Zhi Ma, Cheng Wen, Zhexin Su, Xiao Liang, Cong Tian, Shengchao Qin, and Mengfei Yang.
\newblock Bridging natural language and formal specification--automated translation of software requirements to ltl via hierarchical semantics decomposition using llms.
\newblock {\em arXiv preprint arXiv:2512.17334}, 2025.

\bibitem{Bjorner2000Domain}
Dines Bj{\o}rner.
\newblock Domain engineering: A software engineering discipline in need of research.
\newblock In {\em International Conference on Current Trends in Theory and Practice of Computer Science}, pages 1--17. Springer, 2000.

\bibitem{Wu2024Survey}
Na~Wu, Yifan Li, Hongyang Yang, Hao Chen, Shouyi Dai, and Chenghao Hao.
\newblock Survey of machine learning for software-assisted hardware design verification: Past, present, and prospect.
\newblock {\em ACM Transactions on Design Automation of Electronic Systems}, 2024.

\bibitem{Abdollahi2024HardwareDA}
Meisam Abdollahi, S.~Faegheh Yeganli, Mohammad Baharloo, and Amirali Baniasadi.
\newblock Hardware design and verification with large language models: A scoping review, challenges, and open issues.
\newblock {\em Electronics}, 2024.

\bibitem{Huang2026Modeling}
Yike Huang, Ming Hu, Xiaohong Chen, Zhi Jin, and Shuyuan Xiao.
\newblock Modeling like peeling an onion: Layerwise analysis-driven automatic behavioral model generation.
\newblock In {\em 2026 IEEE/ACM 48th International Conference on Software Engineering (ICSE '26)}, page~12, 2026.

\bibitem{Tang2025VLTLBench}
William English, Chase Walker, Dominic Simon, Sumit~Kumar Jha, and Rickard Ewetz.
\newblock Verifiable natural language to linear temporal logic translation: A benchmark dataset and evaluation suite.
\newblock {\em ArXiv}, abs/2507.00877, 2025.

\bibitem{mendoza2024synttl}
Daniel Mendoza, Christopher Hahn, and Caroline Trippel.
\newblock Translating natural language to temporal logics with large language models and model checkers.
\newblock {\em 2024 Formal Methods in Computer-Aided Design (FMCAD)}, pages 1--11, 2024.

\bibitem{Wang2025ConformalNL2LTL}
Jun Wang, David~Smith Sundarsingh, Jyotirmoy~V. Deshmukh, and Yiannis Kantaros.
\newblock Conformalnl2ltl: Translating natural language instructions into temporal logic formulas with conformal correctness guarantees.
\newblock {\em ArXiv}, abs/2504.21022, 2025.

\end{thebibliography}


\end{document}